\documentclass[12pt,preprint]{aastex}
\usepackage{psfig,amsfonts,amsmath,graphicx,natbib,lscape,nicefrac,url,hyperref,longtable,threeparttable}

\shorttitle{GRB~130427A/SN~2013cq}

\begin{document}
\title{Discovery of the broad-lined Type Ic SN~2013cq associated with the very energetic GRB~130427A}

\author{D.~Xu\altaffilmark{1}, A.~de~Ugarte~Postigo\altaffilmark{2,1}, G.~Leloudas\altaffilmark{3,1}, T.~Kr\"{u}hler\altaffilmark{1}, Z.~Cano\altaffilmark{4}, J.~Hjorth\altaffilmark{1}, D.~Malesani\altaffilmark{1}, J.~P.~U.~Fynbo\altaffilmark{1}, C.~C.~Th\"{o}ne\altaffilmark{2}, R.~S\'anchez-Ram\'irez\altaffilmark{2}, S.~Schulze\altaffilmark{5,6}, P.~Jakobsson\altaffilmark{4}, L.~Kaper\altaffilmark{7}, J.~Sollerman\altaffilmark{8}, D.~J.~Watson\altaffilmark{1}, A.~Cabrera-Lavers\altaffilmark{9}, C.~Cao\altaffilmark{10,11}, S.~Covino\altaffilmark{12}  H. Flores\altaffilmark{13}, 
S.~Geier\altaffilmark{14,1}
J.~Gorosabel\altaffilmark{2,15,16}, S.~M.~Hu\altaffilmark{10}, B.~Milvang-Jensen\altaffilmark{1}, M.~Sparre\altaffilmark{1}, L.~P.~Xin\altaffilmark{17}, T.~M.~Zhang\altaffilmark{17}, W.~K.~Zheng\altaffilmark{18}, Y.~C.~Zou\altaffilmark{19}}

\altaffiltext{1}{Dark Cosmology Centre, Niels Bohr Institute, University of Copenhagen, Juliane Maries Vej 30, 2100 K\o benhavn \O, Denmark; E-mail: dong@dark-cosmology.dk}

\altaffiltext{2}{Instituto de Astrof\'isica de Andaluc\'ia, CSIC, Glorieta de la Astronom\'ia s/n, E - 18008 Granada, Spain}

\altaffiltext{3}{The Oskar Klein Centre, Department of Physics, Stockholm University, AlbaNova, 10691 Stockholm, Sweden}

\altaffiltext{4}{Centre for Astrophysics and Cosmology, Science Institute, University of Iceland, Dunhagi 5, IS-107 Reykjavik, Iceland}

\altaffiltext{5}{Pontificia Universidad Cat\'{o}lica de Chile, Departamento de Astronom\'{\i}a y Astrof\'{\i}sica, Casilla 306, Santiago 22, Chile}

\altaffiltext{6}{Millennium Center for Supernova Science, Chile}

\altaffiltext{7}{Astronomical Institute Anton Pannekoek, University of Amsterdam, Science Park 904, NL-1098 XH Amsterdam, the Netherlands}

\altaffiltext{8}{The Oskar Klein Centre, Department of Astronomy, Stockholm University, AlbaNova, 10691 Stockholm, Sweden}

\altaffiltext{9}{Instituto de Astrof\'isica de Canarias, E-38205 La Laguna, Tenerife, Spain}

\altaffiltext{10}{Department of Space Science and Physics, Shandong University at Weihai, Weihai, Shandong 264209, China}

\altaffiltext{11}{Visiting scholar, Infrared Processing and Analysis Center, Caltech, Pasadena, CA 91125, USA}

\altaffiltext{12}{INAF / Brera Astronomical Observatory, via Bianchi 46, 23807, Merate (LC), Italy}

\altaffiltext{13}{Laboratoire Galaxies Etoiles Physique et Instrumentation, Observatoire de Paris, 5 place Jules Janssen, 92195 Meudon, France}

\altaffiltext{14}{Nordic Optical Telescope, Apartado 474, 38700 Santa Cruz de La Palma, Spain}

\altaffiltext{15}{Unidad Asociada Grupo Ciencia Planetarias UPV/EHU-IAA/CSIC, Departamento de F\'isica Aplicada I, E.T.S. Ingenier\'ia, Universidad del Pa\'is Vasco UPV/EHU, Alameda de Urquijo s/n, E-48013, Bilbao, Spain}

\altaffiltext{16}{Ikerbasque, Basque Foundation for Science, Alameda de Urquijo 36-5, E-48008 Bilbao, Spain}

\altaffiltext{17}{National Astronomical Observatories, Chinese Academy of Sciences, Beijing 100012, China}

\altaffiltext{18}{Department of Astronomy, University of California, Berkeley, CA 94720-3411, USA}

\altaffiltext{19}{School of Physics, Huazhong University of Science and Technology, Wuhan 430074, China}

\shortauthors{Xu et al.}

\begin{abstract}
Long-duration $\gamma$-ray bursts (GRBs) at $z < 1$ are in most cases found to
be accompanied by bright, broad-lined Type Ic supernovae (SNe Ic-BL). The
highest-energy GRBs are mostly located at higher redshifts, where the associated
SNe are hard to detect observationally.
Here we present early and late observations of the optical counterpart of the very energetic GRB~130427A. Despite its moderate redshift $z = 0.3399\pm0.0002$, GRB~130427A is at the high
end of the GRB energy distribution, with an isotropic-equivalent energy release
of $E_{\rm iso} \sim 9.6 \times 10^{53}$~erg, more than an order of magnitude
more energetic than other GRBs with spectroscopically confirmed SNe. In our
dense photometric monitoring, we detect excess flux in the host-subtracted
$r$-band light curve, consistent with that expected from an emerging SN, $\sim
0.2$ mag fainter than the prototypical SN~1998bw. A spectrum obtained around the
time of the SN peak (16.7 days after the GRB) reveals broad undulations typical
of SNe Ic-BL, confirming the presence of a SN, designated SN~2013cq. The
spectral shape and early peak time are similar to those of the high expansion
velocity SN~2010bh associated with GRB~100316D. Our findings demonstrate that
high-energy long-duration GRBs, commonly detected at high redshift, can also be
associated with SNe Ic-BL, pointing to a common progenitor mechanism.
\end{abstract}

\keywords{gamma-ray burst: individual: GRB~130427A --- supernovae: individual: SN~2013cq }

\section{Introduction}
\label{sec:intro}
The standard paradigm for long-duration gamma-ray bursts (GRBs) involves a
broad-lined Type Ic supernova (SN Ic-BL) with $M_V\sim -19$ mag
\citep{Woosley2006,Hjorth2012}, such as those predicted by the collapsar model
\citep{MacFadyen1999}. This is based on spectroscopic evidence in SNe
from low-luminosity GRBs \citep{Bromberg2011}, such as SN~1998bw accompanying
GRB~980425 \citep{Galama1998}, as well as relatively-higher-luminosity GRBs,
such as SN~2003dh accompanying GRB~030329 \citep{Hjorth2003,Stanek2003}.
Interestingly, for the two low-redshift cases of GRB~060505 and GRB~060614 no
associated SN was found to deep limits
\citep{Fynbo2006,DellaValle2006,GalYam2006}, but since then no similar events
have been reported.

\begin{figure*}
\centerline{\includegraphics[width=18cm,height=15cm]{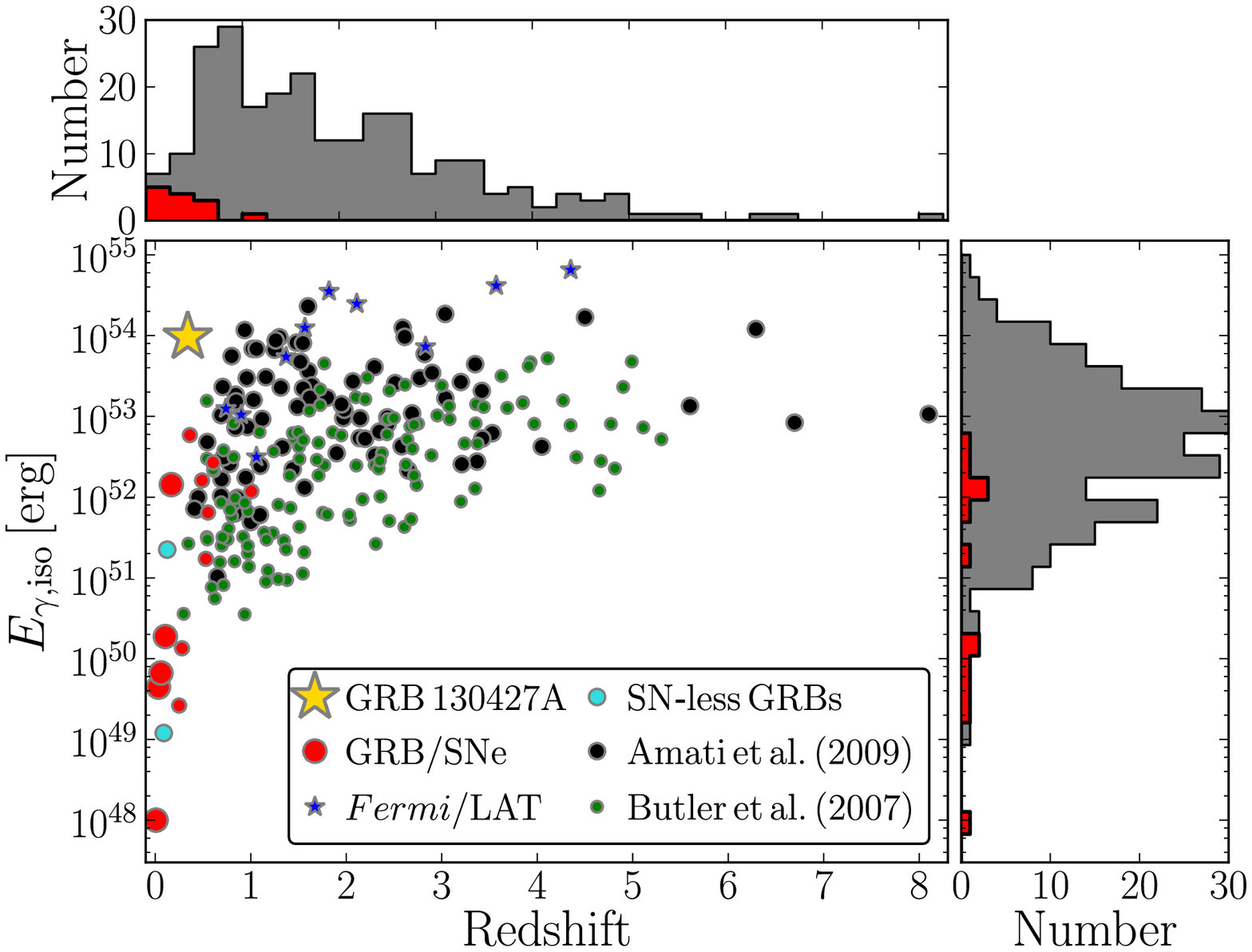}} \caption{\label{eisoz} The isotropic-equivalent energy release in $\gamma$-rays $E_{\rm{iso}}$ of GRB~130427A in comparison to different GRB samples. GRB/SN associations classified with rank A (larger circles) or B (smaller circles) in \citet{Hjorth2012} are plotted in red, and nearby GRBs without a luminous SN are shown with light-blue circles \citep{Fynbo2006}. As blue stars, we show GRBs detected by {\it Fermi}/LAT \citep{McBreen2010,Fermi2013}, in black circles GRBs from \citet{Amati2009} and in green circles {\it Swift} GRBs as derived by \citet{Butler2007}. For GRBs appearing in several of the aforementioned catalogs, earlier referenced data take preference over later one. Due to various reasons such as instrumental constraints, GRB coordinates, weather coordinations and so on, quite a fraction of GRBs at $z < 1$ were not able to be effectively checked whether to be associated with SNe, although they are believed to be with SNe based on those SN-associated GRBs at $z < 1$.}
\end{figure*}

GRB 130427A \citep{Maselli2013,Elenin2013} is remarkable as it is both extremely
energetic and located at a moderately low redshift of $z = 0.3399\pm0.0002$
(\citealt{Levan2013} and this work). Using the spectral
parameters for the prompt emission given by \citet{Kienlin2013}, we derive an
isotropic $\gamma$-ray energy\footnote{We adopt a $\Lambda$CDM cosmology with
$H_0 = 67.3$~km~s$^{-1}$~Mpc$^{-1}$, $\Omega_{\rm m} = 0.315$, and
$\Omega_{\Lambda} = 0.685$ \citep{PC2013}.} of $E_{\rm iso} = (9.61\pm0.04)
\times 10^{53}$~erg in the 1--10000~keV rest-frame energy band. The optical
afterglow peaked at $R \approx 7.4$ mag during the prompt emission phase
\citep{Wren2013}. Only $\sim 5\%$ of all GRBs with measured redshifts are at
such a low distance (see e.g. \citealt{Jakobsson2012}; Figure~\ref{eisoz}), and
those are usually low-luminosity events. By contrast, GRB~130427A was an
extremely energetic burst, and hence it has allowed detailed studies of a system
similar to those usually only found at higher redshifts
\citep{Fan2013,Laskar2013,Tam2013}.

Due to the dearth of such extremely luminous GRBs at low redshift and their
faintness at high redshift, we so far have had no spectroscopic evidence for
accompanying SNe in very energetic GRBs with $E_{\rm iso} > 10^{52.7}$ erg (for photometric evidence of a SN associated with very energetic GRB~080319B, see \citealt{Tanvir2010}). In
Figure~\ref{eisoz} we show the $\gamma$-ray energy release of GRBs as a function
of redshift. Overplotted are systems with spectroscopic evidence for a SN.
GRB~130427A stands out as an exceptional system. It is unclear if progenitor
models involving a SN can power such energetic GRBs (see, e.g.,
\citealt{Piran2004}), and observationally it has proven hard to test the SN
properties of these events. In this work we focus on the search for and
discovery of a SN accompanying this remarkable burst.

\section{Observations and Data Analysis}
\label{sec:obs}
\subsection{Photometry}
Our first follow-up photometry was carried out at the 1-m optical telescope
located in Weihai, Shandong Province, China. The bright optical counterpart of
the GRB was well detected in the Cousins $R_{\rm C}$ filter with $R_{\rm C} \sim 15.5$ mag at 4.178 hr post-burst.
Initially, SDSS filters were
not available at the telescope, but 2 days into our monitoring campaign, SDSS
$r$ and $i$ filters were installed and have been available since then.

\begin{figure*}
\centerline{\includegraphics[width=16cm,height=15cm]{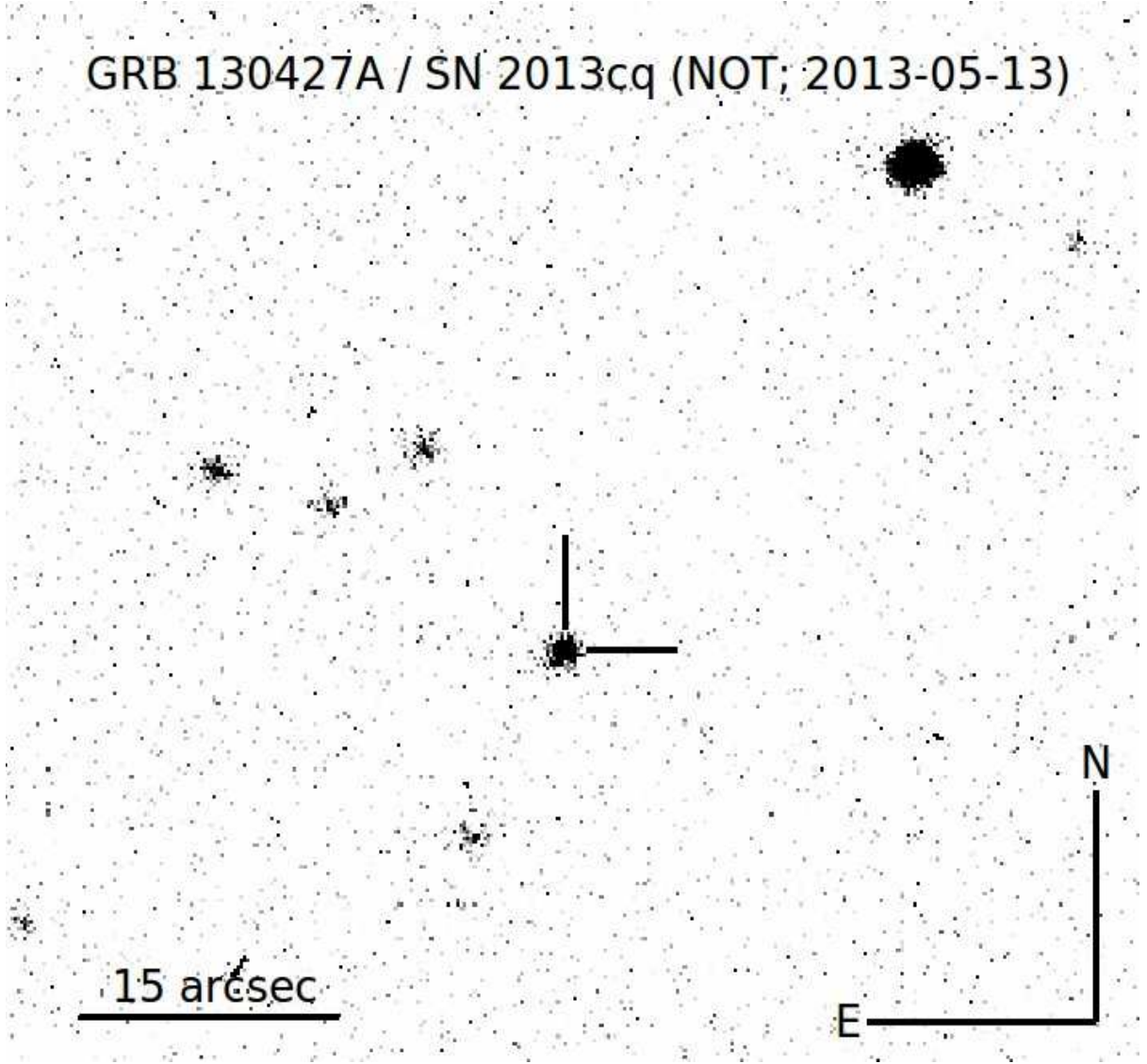}} \caption{\label{field}The field of GRB\,130427A/SN\,2013cq taken at the NOT/ALFOSC at 00:18 UT on 2013-05-13, when it was close to the GTC spectrum time of 00:35 UT on 2013-05-14. North to up and East to left. The angular resolution is 0.19"/pixel and it is indicative that the GRB/SN lies in the northwest part of its extended host galaxy.}
\end{figure*}

Our photometric follow-up observations were mainly obtained at the 2.5-m Nordic
Optical Telescope (NOT) equipped with the ALFOSC instrument. Photometry was
primarily obtained in the $r$-band, complemented by $ugiz$ data useful to
monitor the spectral evolution of the counterpart. As shown in Figure~\ref{field}, there is indication from the NOT
images that the GRB counterpart lies in the NW part of an extended host galaxy.
With our spatial resolution, the afterglow is blended with the host, and we thus use a relatively
big aperture ($\approx 3.8''$ diameter) to measure the magnitudes of the
counterpart plus host. Using a smaller aperture would provide lower fluxes, but
with a host contribution hard to quantify, especially at late times (i.e., since
$\sim 10$ days after the burst). In this way, we consistently include most of
the host light.

Additionally, follow-up observations were obtained at the 2.16-m telescope in
Xinglong, Hebei Province, China. Photometry was done in the Cousins $R_{\rm C}$
and $I_{\rm C}$ filters and then transformed to SDSS $r$ and $i$ magnitudes
based on our afterglow spectra. 

\begin{figure*}
\centerline{\includegraphics[width=14cm,height=20cm,angle=-90]{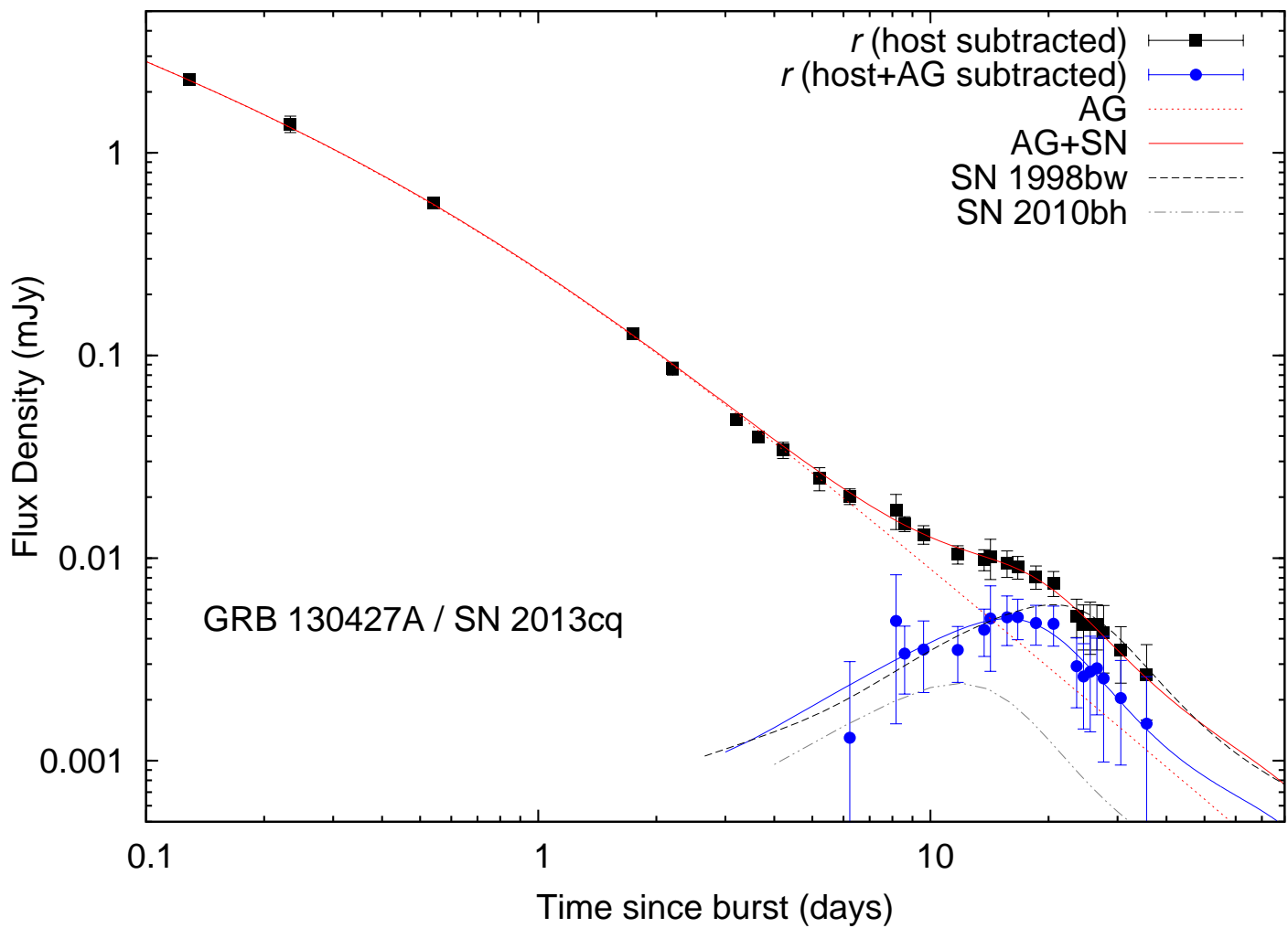}} \caption{\label{rlc}SDSS $r$-band lightcurve of GRB~130427A/SN~2013cq in the observer frame. Filled squares denote host-subtracted magnitudes, while filled circles host- and afterglow- (AG) subtracted magnitudes. The shape and brightness of the latter are consistent with that of a core-collapse supernova. The red dashed line is our AG model (see the text for the best-fitting parameters). The blue solid line plotted against the lightcurve of SN~2013cq is a model supernova. SN~2013cq peaks earlier than SN~1998bw and is about 0.2 mag fainter in the rest-frame $B$ band.}
\end{figure*}

All optical data were reduced in a standard way using IRAF v2.15 in the Scisoft
7.7 package. Magnitudes were calibrated with two nearby bright SDSS field stars, SDSS J113230.55+274420.3 and SDSS J113220.11+274133.5\footnote{\url{http://skyserver.sdss3.org/public/en/tools/chart/navi.aspx?ra=173.1362&dec=27.7129&opt=I}}, whose zero-point errors
are 0.01 mag and propagated into the final magnitude measurements.
The SDSS $r$-band lightcurve is presented in Figure~\ref{rlc} and a log of the
observations is shown in Table~\ref{log}.

We fit the NOT multi-color photometry taken in the first night after the trigger
($g = 17.31\pm0.01$ mag, $r = 17.06\pm0.01$ mag,
$i = 16.92\pm0.01$ mag, $z = 16.86\pm0.02$ mag) and the simultaneous XRT
spectrum\footnote{\url{http://www.swift.ac.uk/xrt_curves/554620}}. Using
synchrotron models and extinction laws from the Local Group \citep[see][for
details]{Kruhler2011}, we estimate the reddening of the host to be $E(B-V)_{\rm
host} = 0.05\pm0.02$~mag for a MW-type extinction law. Within the errors,
this value is consistent with what derived assuming an SMC or LMC extinction
law, because of the small amount of reddening and wavelength range probed by our
observations.

\subsection{Spectroscopy}
Our first spectrum was obtained using NOT/ALFOSC. The total exposure was 1800 s
with a mean time of 0.44 days post-burst. The spectrum covers the range
3200--9100~\AA{} with a resolving power of $\sim 700$. We identify prominent
absorption lines of \ion{Mg}{2} 2796 \& 2803, \ion{Mg}{1} 2852 and
\ion{Ca}{2} 3934 \& 3968, as well as weak emission lines of [\ion{O}{2}] 3727 and H$\beta$, all at a common redshift of $z=0.34$.

A second spectrum with intermediate resolution was obtained shortly afterwards
using the Very Large Telescope (VLT) equipped with the X-shooter spectrograph.
In the spectra the continuum was well detected in the full range
3000--24800~\AA. A number of absorption features are visible, including
\ion{Fe}{2} 2344,
\ion{Mn}{2} 2577,
\ion{Mg}{2} 2796 \& 2803,
\ion{Mg}{1} 2852,
\ion{Ti}{2} 3074,
\ion{Ca}{2} 3934 \& 3968,
\ion{Na}{1} 5890 \& 5896,
and emission lines such as [\ion{O}{2}] 3727, H$\beta$, [\ion{O}{3}] 5007,
and H$\alpha$, all at a common redshift of $z = 0.3399\pm0.0002$.
In the X-shooter spectrum \ion{Na}{1} D 5890 \& 5896 absorption was detected at the redshift
of the host. We measure $0.18 \pm 0.02$ and $0.08 \pm 0.03$~\AA{} for the
equivalent widths of the \ion{Na}{1} D1 and D2 components. Using the relations
in \citet{Poznanski2012}, we obtain an estimate for $E(B-V)_{\rm host} = 0.03
\pm 0.01$~mag, but remark that there exists substantial dispersion of  $E(B-V)
\sim 0.15$~mag in this relation. Considering different calibrations/systematics
involved in above $E(B-V)_{\rm host}$ measurements, we adopt $E(B-V)_{\rm
host}=0.05$\,mag for the host extinction.

Given the relatively low redshift, we planned a third spectroscopic observation
with the aim of detecting the SN signatures. Based on the light curve evolution,
we obtained a spectrum of the optical counterpart and host galaxy with the
10.4-m Gran Telescopio Canarias (GTC) 16.7 days after the GRB. This corresponds
to 12.5 days in the GRB rest frame. Observations consisted of $4\times1200$~s,
covering the range of 4800--10000~\AA{} with a resolving power of $\sim 600$.
The slit was oriented to cover both the afterglow position and the host galaxy
nucleus.

All spectroscopic data were reduced in a standard way using dedicated IRAF pipelines or ESO pipelines. The resulting spectra are presented in Figure~\ref{spectrum}. 

\subsection{Host Galaxy}

We use the catalogued pre-explosion imaging from the SDSS Data Release \citep{Aihara2011}
to estimate physical properties of the
GRB host galaxy and to build a physical model of the stellar emission. The SDSS
$ugriz$ photometry was fitted within LePhare \citep{Arnouts1999,Ilbert2006} 
using stellar population synthesis models from \citet{Bruzual2003} as detailed
in \citet{Kruhler2011}. Based on the model fit to the data
(Figure~\ref{spectrum}), we derive a luminosity $M_B = -19.8 \pm 0.2$ mag, a
stellar mass of $M_\star = 10^{9.0 \pm 0.2}\,M_{\sun}$, a star-formation rate
${\rm SFR_{SED}} = 2_{-1}^{+5}\,M_{\sun}$~yr$^{-1}$ and an age of the starburst
of $\tau = 400_{-250}^{+560}$~Myr for the host of GRB~130427A.

Host galaxy emission lines are detected above the SN continuum, including
[\ion{N}{2}] 6584 in the GTC spectrum, albeit at low significance. This allows us
to place constraints on the metallicity of the explosion host environment by
using the calibrations in \citet{Pettini2004}. We measure $\log{(\rm{O/H})} + 12
= 8.43 \pm 0.07$ and $8.51 \pm 0.09$ by using the O3N2 and the N2 methods,
respectively (statistical errors only), which is at the top-right of the GRB-SN
range in the metallicity$-M_{B,{\rm host}}$ plane and similar to the cases of SNe Ic-BL without observed GRBs and SNe Ib+IIb (see Fig.~2 in
\citealt{Modjaz2011}). We note that the host galaxy also nicely lies within the 1$\sigma$ dispersion of the mass-metallicity relation for normal field galaxies \citep{Kewley2008}. After including the systematic dispersion of 0.14 and
0.18 dex \citep{Pettini2004} for the two methods, these results translate to
a metallicity of $0.55 \pm 0.19$ and $0.67 \pm 0.25$ $Z_{\odot}$, respectively
\citep{Asplund2009}.

\section{SN~2013cq Associated with GRB~130427A}
\label{sec:supernova}

\subsection{Decomposition of the GTC Spectrum}
\label{subsec:gtc}

\begin{figure*}
\centerline{\includegraphics[width=18cm,height=18cm]{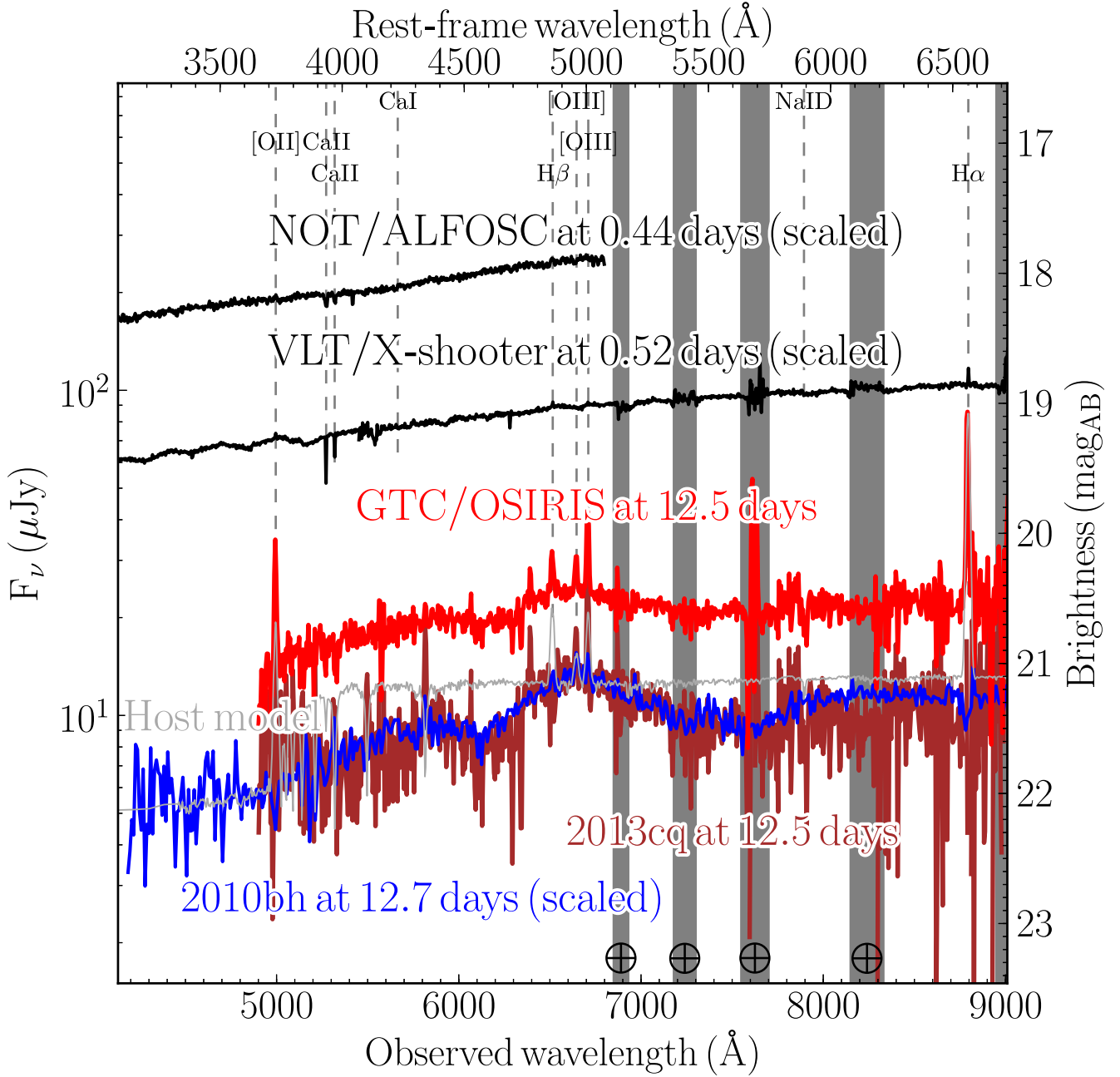}} \caption{\label{spectrum} Spectroscopic sequence of GRB~130427A/SN~2013cq. Epochs of the spectra in their rest frames are marked, together with prominent emission and absorption lines and telluric bands (circled crosses as well as grey shaded areas). The top two (black) were obtained from NOT/ALFOSC and VLT/X-shooter in the burst night. The
red part of the NOT spectrum is affected by fringing (in grey). The original GTC, the host galaxy, and the host-subtracted GTC spectra are in red, grey, and wine, respectively. The host-subtracted spectrum is well matched by that of broad-lined Type Ic SN~2010bh at 12.7 days post-burst, very close to 12.5 days of SN~2013cq here.}
\end{figure*}

We scaled the GTC spectrum with our simultaneous photometry from the same night,
dereddened it for $E(B-V)_{\rm MW}=0.02$\,mag \citep{Schlegel1998} in the Milky Way galaxy, 
and then subtracted the model host galaxy spectrum, after bringing it to the
same resolution, in order to obtain a ``clean'' spectrum of the transient. Afterwards, the spectrum
is dereddened by $E(B-V)_{\rm host}=0.05$\,mag for the host extinction at $z = 0.34$.
Both the original and the final GTC spectra are shown in Figure~\ref{spectrum}.

Although the resulting spectrum is noisy, it shows clear SN features, with the
most prominent being a strong bump peaking at $\sim 6700$~\AA{} (observer-frame;
$\sim 5000$~\AA{} rest-frame). The features are broad (and no H or He can be
seen), justifying the classification of SN~2013cq as a SN Ic-BL
\citep{deUgartePostigo2013}.

SN~1998bw \citep{Patat2001}, associated with GRB~980425, does not provide a good
spectral match to SN~2013cq, mainly because its main peak is located more to the
red (rest-frame $\sim 5200$~\AA{} at similar phases). The same is true for
SN~2006aj \citep{Pian2006,Sollerman2006}, associated with GRB~060218. Instead,
we find a better match with SN~2010bh, associated with GRB~100316D, which is
known to have high expansion velocities, up to 10000 km s$^{-1}$ higher than
other previous SNe Ic-BL associated with GRBs at all phases \citep{Bufano2012}.
In particular, the best match is obtained with the spectrum of SN~2010bh at
a rest-frame time of 12.7 days, very close to the rest-frame 12.5 days for
SN~2013cq here (shown in Figure~\ref{spectrum}).
The similarity to SN\,2010bh is striking although there may be small colour differences (e.g., difference redward of $\sim$7500~\AA{}), which may reflect the diversity in GRB-SN spectra and/or the uncertain extinction correction towards SN\,2010bh.
Considering that the P-Cygni feature on the left of the strong bump can be
primarily attributed to \ion{Fe}{2} 5169, we measure an expansion velocity of
$v_{\rm ph} \sim 32000$ km s$^{-1}$ from the absorption minimum. This is very
similar to the peak photospheric velocity, $v_{\rm peak} \sim 35000$ km
s$^{-1}$, of SN~2010bh \citep{Bufano2012}.

\subsection{Decomposition of the $r$-band Lightcurve}
\label{subsec:rlc}
The $r$-band lightcurve presented in Figure~\ref{rlc} has been corrected for the
foreground extinction ($E(B-V)_{\rm MW}=0.02$ mag). From
the foreground-extinction-corrected flux we subtract the contribution of the
host galaxy ($r_{\rm host} = 21.26 \pm 0.09$, as determined via photometry
on pre-explosion SDSS imaging), so that the flux powering the lightcurve can be
attributed solely to the afterglow and accompanying SN. Visual inspection of
the light curve reveals a deviation away from a power-law decay a few days
post-burst, followed by a plateau that lasts for about 10 days, before decaying
further. At the GTC spectrum time the Galactic-extinction-corrected flux of the
SN/afterglow is 1.85 times that of the host galaxy.

Next, we fit a smoothly broken power-law model \citep{Beuermann1999} to the
lightcurve up to the first four days (some $r$-band data points $<3$~days
published in GCNs are used for the fitting;
\citealt{Perley2013,Wiggins2013,Butler2013,Zhao2013}
), and derive the following
best-fit parameters: $\alpha_1 = 0.69 \pm 0.13$, $\alpha_2 = 1.66 \pm 0.18$, and
$t_{\rm break} = 0.62 \pm 0.48$ days (the smoothness factor of the break is fixed to be unity and $\chi^2$/dof = 1.47 for 7 dof). This afterglow model is then subtracted
from the already host-subtracted lightcurve, and the resultant flux resembles
that of a supernova both in brightness and shape (e.g., \citealt{Cano2011a}).
The peak brightness of SN~2013cq is found to be $r = 22.13$ mag at $\sim 15.2$ days
observer-frame ($\sim 11.3$ days rest-frame). At $z=0.34$, the observer-frame
$r$ band corresponds approximately to the rest-frame $B$ band, and
we find that SN~2013cq has an absolute $B$-band magnitude of $M_B = -18.97 \pm
0.14$, which is about 0.2 mag fainter than SN~1998bw.

We then compare the optical properties of SN~2013cq relative to two other
GRB-associated SNe: SN~1998bw and SN~2010bh. We create synthetic $r$-band
light curves of SN~1998bw and SN~2010bh as they would appear at $z=0.34$.
Next, using Equations 1 and 2 from \citet{Cano2011b}, we determine the stretch
($s$) and luminosity factor ($k$) of SN~2013cq relative to SN~1998bw to be
$s = 0.77\pm 0.03$ and $k = 0.85 \pm 0.03$.

We further estimate the bolometric properties of SN~2013cq. The method,
developed by \citet{Cano2013}, uses the relative stretch and luminosity factors
of a SN relative to a template (in this case SN~1998bw), and makes the
assumption that the relative shape of a given SN in a given filter is a good
proxy for the relative shape of the SN bolometric light curve relative to the
template. The bolometric light curve of the template\footnote{Note that we are
using and transforming a bolometric light curve of SN~1998bw that has been constructed
using its observational data in $UBVRIJH$ filters.} is then altered by $s$ and
$k$, and then fit with an analytical model derived from \citet{Arnett1982}. The
Arnett model depends on knowing the photospheric velocity at peak bolometric
light to determine the ejecta mass and explosion energy, which for many SNe is
determined from the velocity of different species in the SN spectra at peak
light.  We have used the photospheric velocity value determined in Section
\ref{subsec:gtc} (i.e. $v_{\rm ph} \sim 32000$ km s$^{-1}$). On comparison with SN~1998bw,
for SN~2013cq (in $UBVRIJH$) we derived its Ni mass, ejecta mass, and kinetic
energy $M_{\rm Ni} = 0.28 \pm 0.02 \, M_{\odot}$, $M_{\rm ej} = 6.27 \pm
0.69 \, M_{\odot}$, and $E_{\rm K} = (6.39 \pm 0.70) \times 10^{52}$ erg,
respectively. The quoted errors are statistical only.

\section{Discussion}
\label{sec:disc}

Our photometric and spectroscopic campaign has led to the unambiguous discovery
of a SN Ic-BL, SN~2013cq \citep{deUgartePostigo2013}, associated with GRB~130427A at $z=0.34$. GRB~130427A is one of the most energetic bursts ever
detected with $E_{\rm iso} \sim 9.6 \times 10^{53}\,{\rm erg}$, comparable to
that of high-redshift GRBs and much larger than local events. The fact that a
supernova progenitor model accounts for even very energetic bursts suggests a
common progenitor model, such as the collapsar model, may account for the
majority of all long-duration GRBs. To overcome the challenge of providing
enough energy to power the GRB, it is likely that the large observed energy is
due to beaming (for a strong beaming of GRB~130427A, see
\citealt{Laskar2013}), making the true energy much lower (typically two orders
of magnitude).

Our discovery now suggests that not only core-collapse SNe, but specifically   
stripped envelope, high velocity SNe are almost an inevitable consequence of
the deaths of stars that form all GRBs. A common mechanism is therefore at
play to power GRBs with very different high-energy properties.

It is worth noting that the comparable peak $B$-band luminosity of SN~2013cq and
of SN~1998bw is consistent with the suggestion of \citet{Hjorth2013}
that there is an upper envelope to the brightness of GRB-SNe which drops
slightly with isotropic luminosity. The origin of such an upper envelope is
intriguing, but currently not clear.

As shown in \citet{Modjaz2011}, among previous GRB-SN events, the highest oxygen abundance $\log{(\rm{O/H})} +
12$ at the SN position was less than 8.3, and when the oxygen
abundance rises towards a seemingly critical value of 8.5, GRB-SN events tend to locate
in dwarf galaxies with $M_{B,{\rm host}} > -19.0$ mag. Note that the abundance value of 8.5 is a typical one for SNe Ic-BL without observed GRBs which occur in both luminous and dwarf galaxies and also it is typical for SNe Ib+IIb which happen in relatively luminous galaxies with $M_{B,{\rm host}} < -18.5$ mag \citep{Modjaz2011}. With an abundance of $8.43  \pm 0.07$ for the SN position and $M_B = -19.8 \pm 0.2$ mag for the host, GRB~130427A/SN~2013cq is consistent with subclasses of core-collapse SNe such as SNe Ic-BL without observed GRBs and SNe Ib+IIb in the metallicity$-M_{B,{\rm host}}$ plane, implying that GRBs do not exclusively explode in low metallicity dwarf galaxies.

\acknowledgements
We thank the referee for valuable comments and suggestions. We are grateful to Jens Jessen-Hansen, Tiina Liimets, Dagmara Oszkiewicz, Yeisson Martinez-Osorio, Olesja Smirnova, Jari Kajava, Tuomas Kangas, Emanuel Gafton, Anthony Macchiavello, Tore Mansson, and Joachim Wiegert for taking images at the NOT. We are also grateful to Jinyi Yang for carrying out observations at the Xinglong 2.16-m telescope. DX thanks Adam S. Trotter for
helpful discussion. DX, JPUF and BMJ acknowledge support from
the ERC-StG grant EGGS-278202. AdUP acknowledges support by the European Commission under
the Marie Curie Career Integration Grant programme (FP7-PEOPLE-2012-CIG 322307).
AdUP, CT and JG are supported by the Spanish research project
AYA2012-39362-C02-02. GL is supported by the Swedish Research Council through
grant No. 623-2011-7117. ZC and PJ acknowledge support from the Icelandic
Research Fund. SS acknowledges financial support from the Iniciativa Cientifica
Milenio grant P10-064-F (Millennium Center for Supernova Science), with input
from "Fondo de Innovaci\'{o}n para la Competitividad, del Ministerio de
Econom\'{\i}a, Fomento y Turismo de Chile", and Basal-CATA (PFB-06/2007).
LPX, TMZ, and YCZ acknowledge the support of the National Science Foundation of China
grants 11103036, 11203034 \& 11178003, and U1231101, respectively.
The Dark Cosmology Centre is funded
by the Danish National Research Foundation. 
Based on observations made with the Nordic Optical Telescope, operated by the
Nordic Optical Telescope Scientific Association at the Observatorio del Roque de
los Muchachos, La Palma, Spain, of the Instituto de Astrof\'isica de Canarias.
Based on observations made with ESO Telescopes at the La Silla Paranal Observatory under programme ID 091.C-0934
Based on observations made with the Gran Telescopio Canarias (GTC), at the Spanish
Observatorio del Roque de los Muchachos (La Palma). 
This work made use of data supplied by the UK {\it Swift} Science Data Centre at the University of Leicester.

\clearpage

\clearpage

\clearpage

\clearpage

\clearpage
\begin{table}
\label{log}
\centering
\caption{Log of the SDSS $r$-band observations and photometry.}
\begin{tabular}{llll}
\hline \hline
Mid time (days)\tablenotemark{a} & Mag\tablenotemark{b} & Error & Instrument \\ \hline
0.538 & 17.06 & 0.02 & NOT/ALFOSC \\
1.745 & 18.59 & 0.02 & NOT/ALFOSC \\
2.204 & 18.98 & 0.07 & WH/PI \\ 
3.205 & 19.51 & 0.05 & WH/PI \\
3.648 & 19.68 & 0.02 & NOT/ALFOSC \\
4.21 & 19.80 & 0.07 & WH/PI \\ 
5.21 & 20.05 & 0.09 & WH/PI \\
6.23 & 20.19 & 0.05 & XL/BFOSC \\
8.17 & 20.30 & 0.12 & WH/PI \\  
8.60 & 20.39 & 0.03 & NOT/ALFOSC \\ 
9.60 & 20.46 & 0.04 & NOT/ALFOSC \\ 
11.74 & 20.58 & 0.02 & NOT/ALFOSC \\  
13.71 & 20.61 & 0.03 & NOT/ALFOSC \\  
14.22 & 20.60 & 0.10 & XL/BFOSC \\  
15.69 & 20.63 & 0.05 & NOT/ALFOSC \\  
16.70 & 20.65 & 0.03 & GTC/OSIRIS\\  
18.59 & 20.70 & 0.02 & NOT/ALFOSC \\  
20.59 & 20.73 & 0.02 & NOT/ALFOSC \\  
23.61 & 20.87 & 0.03 & NOT/ALFOSC \\
24.57 & 20.90 & 0.04 & NOT/ALFOSC\\
25.56 & 20.90 & 0.06 & NOT/ALFOSC\\
26.57 & 20.90 & 0.04 & NOT/ALFOSC\\
27.64 & 20.93 & 0.08 & NOT/ALFOSC\\
30.58 & 20.98 & 0.03 & NOT/ALFOSC\\
35.57 & 21.04 & 0.03 & NOT/ALFOSC\\\hline 
\end{tabular} 
\tablenotetext{a} {The middle time of each epoch, in units of day post-burst, is relative to the {\it Fermi} trigger time of 07:47:06 UT on 2013-04-27, $\sim 51$ s ahead of the {\it Swift} trigger time of this burst.}
\tablenotetext{b} {Magnitudes not corrected for Galactic reddening of $E(B-V)_{\rm MW}=0.02$ mag.}
\end{table}

\end{document}